\RequirePackage{fix-cm}
\documentclass[smallextended]{svjour3}       
\smartqed  
\usepackage{graphicx}
\usepackage{multirow}
\usepackage{amsmath}
\usepackage{mathrsfs}
\usepackage{amsfonts}
\usepackage{amssymb}
\usepackage{booktabs}
\begin{document}

\title{Efficient Fusion of Photonic $W$-states with Nonunitary Partial-swap Gates}


\author{Hai-Rui Wei$^{1*}$,  Wen-Qiang Liu$^1$   and   Leong-Chuan Kwek$^{2,3,4}$ }


\institute{Hai-Rui Wei \at
                $^1$ School of Mathematics and Physics, University of Science and Technology Beijing, Beijing 100083, China\\
            Wen-Qiang Liu \at
                 $^1$ School of Mathematics and Physics, University of Science and Technology Beijing, Beijing 100083, China\\
          Leong-Chuan Kwek \at
               $^2$Centre for Quantum Technologies, National University of Singapore, Singapore 117543, Singapore \\
               $^3$ MajuLab, CNRS-UNS-NUS-NTU International Joint Research Unit,  Singapore UMI 3654, Singapore\\
               $^4$ National Institute of Education and Institute of Advanced Studies, Nanyang Technological University, Singapore 637616, Singapore\\
               \email{hrwei@ustb.edu.cn}
}

\date{Received: date / Accepted: date}

\maketitle

\begin{abstract}

We introduce a nonunitary partial-swap gate for fusing arbitrary  small-sized photonic $W$-states into a large-scale entangled network of $W$-state efficiently without ancillary photons. A partial-swap gate is designed in an optical architecture based on linear optics elements. By introducing auxiliary degree of freedom, this gate provides a higher success probability with less cost.  Our implementation can create a larger target state with a simpler set-up than previous proposals for $W$-state fusion. Also all ``garbage'' states are recyclable, i.e., there is no complete failure output in our scheme in principle.

\end{abstract}

\vspace{2pc}

\section{Introduction}\label{sec1}

Quantum entanglement is a key resource in many quantum information processing (QIP) tasks \cite{book}, such as quantum teleportation \cite{teleportation1,teleportation2}, quantum superdense coding \cite{superdense}, quantum key distribution  \cite{distribution}, quantum algorithms \cite{algorithm}, and measurement-based quantum computation \cite{one-way}.  In particular, depending on the requirement of a specific task, the preparation and the manipulation of multiqubit entangled states with different multipartite features (for instance, the Greenberger--Horne--Zeilinger  \cite{GHZ}, Dicke \cite{Dick}, $W$ \cite{W}, and cluster states \cite{cluster}) are needed, and these different types of entangled states cannot be converted into each other by using local operations and classical communications.  However, there are still theoretical and experimental challenges in study of multipartite entanglement due to the more complex mathematical form and rapidly growing resource overhead with the number of particles increasing.
In recent years,  the weblike property of the $W$-state, due to its robustness against particle loss and decoherence effects, has rendered it to be a useful resource for quantum communication \cite{fujii2011robust}.  Indeed, the $W$-state has been shown to be relevant for many schemes and applications ranging from its use in the foundation of quantum mechanics \cite{test,mattar2017experimental,wu2014robust}, in anonymous quantum networks \cite{network}, in quantum telecloning and teleportation \cite{telecloning},  in quantum computation \cite{cloning-machine,Ai}, in quantum memories \cite{choi2010entanglement} and as a probe for reading information \cite{guha2013reading}.


So far, many theoretical proposals  and realistic experiments  for generating small-size $W$ states have been proposed \cite{exp-W1,exp-W2,exp-W3}.
Currently, there are two efficient ways to generate large-scale photonic  $W$ states: expansion and fusion. Both schemes have now been achieved in a wide range of physical settings \cite{optics1,optics2,ion,NMR}. In 2008, Tashima \emph{et al}. \cite{optics1} proposed a scheme for locally expanding any polarization-based $n$-photon $W$ ($|W_n\rangle$) state to an $(n+2)$-photon $W$ ($|W_{n+2}\rangle$) state by accessing just one of the qubits with a success probability of $(n+2)/(16n)$. This scheme was subsequently  demonstrated experimentally in 2010 \cite{demonstrated}. Schemes for  expanding $|W_n\rangle$ locally to $|W_{n+1}\rangle$ with a success probability of $(n+1)/(5n)$ were also proposed in 2009 \cite{expand2009}, and even one for $|W_n\rangle$ to $|W_{n+k}\rangle$ was proposed in 2011 \cite{expand2011}. Notably, the success probability of the expansion from $|W\rangle_n$ to $|W_{n+k}\rangle$ decreases with an approximately exponential dependence with increasing $n$.


Fusion, on the other hand, was first proposed in 2011 by \"{O}zdemir \emph{et al}. \cite{Ozdemir}.  The idea was to join $|W_n\rangle$ and $|W_m\rangle$ to give the $|W_{n+m-2}\rangle$ with a success probability of $(n+m-2)/(nm)$.
In 2013, enhancement the $W$-state fusion process was proposed through the use of a Fredkin gate, Bugu \emph{et al}. \cite{Bugu} then proposed a mechanism to fuse $|W_n\rangle$ and $|W_m\rangle$ with one ancillary single photon into $|W_{n+m-1}\rangle$ with a success probability of $(n+m-1)/(mn)$. In 2014, Ozaydin \emph{et al}. \cite{Ozaydin} generalized the setup for fusing three $W$ states: $|W_n\rangle$, $|W_m\rangle$, and $|W_t\rangle$ states and one ancillary single photon were joined  into $|W_{n+m+t-3}\rangle$ with a success probability of $(n+m+t-3)/(mnt)$ with a Fredkin gate. Using three CNOT gates and one Toffoli gate, Yesilyurt \emph{et al}. \cite{Yesilyurt} further generalized the scheme for creating $|W_{n+m+t+z-4}\rangle$ from $|W_n\rangle$, $|W_m\rangle$, $|W_z\rangle$, and $|W_t\rangle$ states with a success probability of $(n+m+t+z-4)/(mntz)$. However, the success probabilities of the required CNOT \cite{KLM,1/9,1/41,1/42},  Toffoli \cite{Toffoli0,Toffoli1,Toffoli,Toffoli2,Toffoli3}, and Fredkin \cite{Fredkin1,Fredkin,Fredkin2,Fredkin3,Fredkin4} gates with linear optical elements were generally not considered. Currently, nonlinear fusion schemes for fusing $|W_n\rangle$ and $|W_m\rangle$ into $|W_{n+m}\rangle$ without qubit loss have also been proposed \cite{loss,Gaoting}.


In this paper, we propose a protocol for fusing $W$ states of arbitrary size into a larger one via nonunitary partial-swap gates. By introducing auxiliary spatial degrees of freedom and using $(n-1)$ partial-swap gates, a $|W_{nm-n+1}\rangle$ state can be created from $n$ arbitrary-sized $|W_m\rangle$ states. All the ``garbage''  states are recyclable, and our scheme avoids failed outcomes. Moreover, additional ancillary photon is not required for our scheme. The length (cost or complexity) of our scheme (measured by the number of the two-qubit entangling gates needed to construct the scheme) is much less than the Fredkin- and CNOT-Toffoli-based schemes \cite{Bugu,Yesilyurt}.

\begin{figure}[htpb]
\begin{center}
\includegraphics[width=7 cm,angle=0]{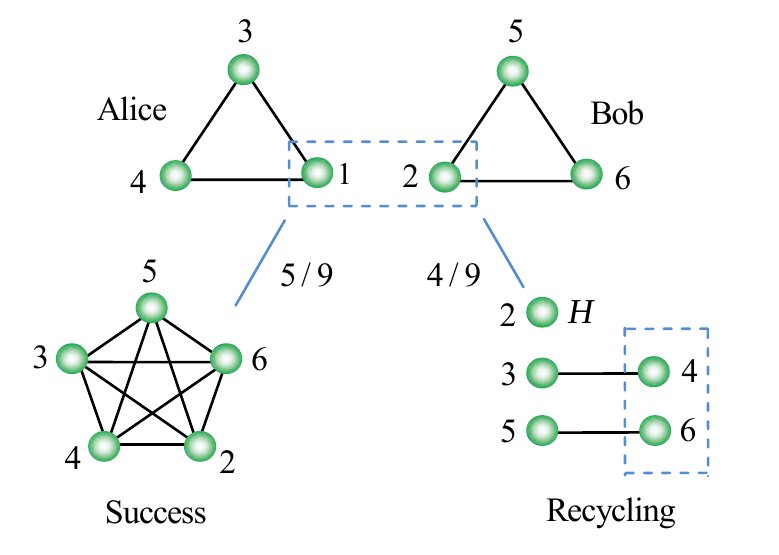}
\caption{Schematic diagram of the proposed scheme for fusing two $|W_3\rangle$ states into a larger one $|W_5\rangle$. The fusion gate operates on the two qubits in the dashed blue rectangle.}  \label{Fusion}
\end{center}
\end{figure}

\section{Simplifying a fusion-based $W$ state with nonunitary partial-swap gate} \label{sec2}

\subsection{Fusion of $|W_n\rangle$ and $|W_m\rangle$ to give $|W_{n+m-1}\rangle$ }

%

Suppose Alice and Bob possess $n$- and $m$-partite polarization encoded $W$ states, $|W_n\rangle_A$ and $|W_m\rangle_B$, respectively, and they wish to fuse their states together. A schematic example for the fusion  process of two three-partite $W$-states is depicted in Fig. \ref{Fusion}. The entangled $W$-states of Alice ($|W_n\rangle_A$)  and Bob ($|W_m\rangle_B$) can be written as
\begin{eqnarray}              \label{eq1}
\begin{split}
|W_n\rangle_A=(|(n-1)_H\rangle_a|1_V\rangle_1+\sqrt{n-1}|W_{n-1}\rangle_a|1_H\rangle_1)/\sqrt{n},
\end{split}
\end{eqnarray}
\begin{eqnarray}              \label{eq2}
\begin{split}
|W_m\rangle_B=(|(m-1)_H\rangle_b|1_V\rangle_2+\sqrt{m-1}|W_{m-1}\rangle_b|1_H\rangle_2)/\sqrt{m},
\end{split}
\end{eqnarray}
where $|(N-k)_H\rangle_i|k_V\rangle_j$ represents the superposition of all possible permutations of $N-k$ photons with a horizontal polarization ($H$) in mode $i$ and $k$ photons with a vertical polarization ($V$) in mode $j$. Captial letters $A$ and $B$ label the $W$ states are held by Alice and Bob, respectively. Therefore, the initial state of the system composed of Alice and Bob can be written as
\begin{eqnarray}              \label{eq3}
\begin{split}
|W_n\rangle_A\otimes|W_m\rangle_B=
&\frac{1}{\sqrt{nm}}                |(n-1)_H\rangle_a|(m-1)_H\rangle_b|1_V\rangle_1|1_V\rangle_2\\
&+\frac{\sqrt{m-1}}{\sqrt{nm}}       |(n-1)_H\rangle_a|W_{m-1}\rangle_b|1_V\rangle_1|1_H\rangle_2\\
&+\frac{\sqrt{n-1}}{\sqrt{nm}}       |W_{n-1}\rangle_a|(m-1)_H\rangle_b|1_H\rangle_1|1_V\rangle_2\\
&+\frac{\sqrt{(n-1)(m-1)}}{\sqrt{nm}}|W_{n-1}\rangle_a|W_{m-1}\rangle_b|1_H\rangle_1|1_H\rangle_2.
\end{split}
\end{eqnarray}

\begin{figure}[htpb]
\begin{center}
\includegraphics[width=8 cm,angle=0]{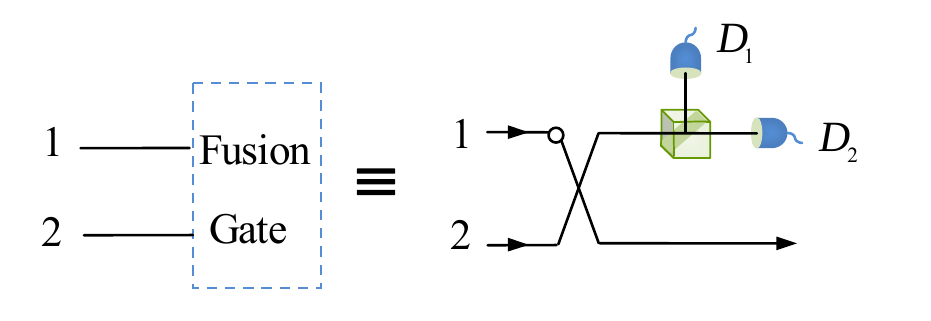}
\caption{Fusion gate for fusing two $W$ states of arbitrary size to obtain a larger $W$ state. The circle ``$\circ$'' denotes $|H\rangle$, signifying that the states of the two photons in modes 1 and 2 will be exchanged if the photon in mode 1 is the $H$-polarized states, and has no effect otherwise. $D_1$ and $D_2$ are single-photon detectors.} \label{w2}
\end{center}
\end{figure}



\begin{table}[htb]
\centering\caption{Truth table of the polarization partial-swap gate.}
\begin{tabular}{ccccc}

 \toprule

 $x_1x_2$      &   $\rightarrow$    &  $y_1y_2$  &        \\

\midrule

  $H_1H_2$          &        &  $H_1H_2$      \\
  $H_1V_2$          &        &  $V_1H_2$      \\
  $V_1H_2$          &        &  $V_1H_2$      \\
  $V_1V_2$          &        &  $V_1V_2$      \\

\bottomrule
\end{tabular}\label{Truth}
\end{table}


As shown in Fig. \ref{w2}, the fusion of $|W_m\rangle_A$ and $|W_n\rangle_B$ states into a larger $W$ state is achieved by sending photons in  mode 1 (2), i.e., $|1_H\rangle_1$ and $|1_V\rangle_1$ ($|1_H\rangle_2$ and $|1_V\rangle_2$), into the partial-swap gate and those in  mode a (b) are kept intact at Alice's (Bob's) site.  Note that the partial-swap gate swaps the states of the two photons if the first photon is $H$ polarization, and has no effect otherwise (see Table \ref{Truth}). That is, the action of this gate on the four input states yields
\begin{eqnarray}              \label{eq4}
\begin{split}
&|1_H\rangle_1|1_H\rangle_2 \xrightarrow{\text{p-swap}} |1_H\rangle_1|1_H\rangle_2,\quad|1_H\rangle_1|1_V\rangle_2 \xrightarrow{\text{p-swap}} |1_V\rangle_1|1_H\rangle_2,\\
&|1_V\rangle_1|1_H\rangle_2 \xrightarrow{\text{p-swap}} |1_V\rangle_1|1_H\rangle_2,\quad|1_V\rangle_1|1_V\rangle_2 \xrightarrow{\text{p-swap}} |1_V\rangle_1|1_V\rangle_2.
\end{split}
\end{eqnarray}
Therefore, such partial-swap gate completes the transformation
\begin{eqnarray}              \label{eq5}
\begin{split}
|W_n\rangle_A\otimes|W_m\rangle_B\rightarrow
&\frac{1}{\sqrt{nm}}                |(n-1)_H\rangle_a|(m-1)_H\rangle_b|1_V\rangle_1|1_V\rangle_2\\
&+\frac{\sqrt{m-1}}{\sqrt{nm}}       |(n-1)_H\rangle_a|W_{m-1}\rangle_b|1_V\rangle_1|1_H\rangle_2\\
&+\frac{\sqrt{n-1}}{\sqrt{nm}}       |W_{n-1}\rangle_a|(m-1)_H\rangle_b|1_V\rangle_1|1_H\rangle_2\\
&+\frac{\sqrt{(n-1)(m-1)}}{\sqrt{nm}}|W_{n-1}\rangle_a|W_{m-1}\rangle_b|1_H\rangle_1|1_H\rangle_2\\
=\;&\frac{\sqrt{n+m-1}}{\sqrt{nm}}|W_{n+m-1}\rangle_{a,b,2}|1_V\rangle_1+\frac{\sqrt{(n-1)(m-1)}}{\sqrt{nm}}\\&\times|W_{n-1}\rangle_a|W_{m-1}\rangle_b|1_H\rangle_1|1_H\rangle_2.
\end{split}
\end{eqnarray}
The photon in mode 1 is then measured in the $\{|H\rangle,\; |V\rangle\}$ basis.  From Eq. (\ref{eq5}), one  sees that (i) When the photon in mode 1 is $V$-polarized and detector $D_1$ clicks, the scheme is successful with probability (success probability) of $(n+m-1)/(nm)$, and the system collapses into the desired state
\begin{eqnarray}              \label{eq6}
\begin{split}
\frac{\sqrt{n+m-1}}{\sqrt{nm}}|W_{n+m-1}\rangle_{a,b,2}.
\end{split}
\end{eqnarray}
(ii) When the photon in mode 1 is $H$-polarized state and detector $D_2$ clicks, then the remaining photon collapses into state  $|W_{n-1}\rangle_a\otimes|W_{m-1}\rangle_b\otimes|1_H\rangle_2$ with probability (recycle probability) of $(n-1)(m-1)/(nm)$.  It is interesting to see that the ``garbage'' state $|W_{n-1}\rangle_a\otimes|W_{m-1}\rangle_b$ of Alice and Bob remains a $W$ state but with a reduced number of qubits, and therefore this state can be recycled, much like a repeat-until-success scheme \cite{lim2005repeat}. Remarkably, the fail probability of the designed scheme is zero in principle as the system can not collapse into the failure states, such as $|(n-1)_H\rangle_a|(m-1)_H\rangle_b|1_V\rangle_2$.


In Table \ref{table1}, we compare our scheme with previous protocols. Here the success probabilities of the Fredkin, Toffoli, CNOT, and partial-swap gates are disregarded. The linear optical entangling gates are inherently probabilistic. The optimal cost of a Fredkin or Toffoli gate is five two-qubit gates \cite{Optimal1,Optimal2}, therefore, the complexity of our partial-swap-based scheme is much lower than the Fredkin-based one \cite{Bugu} and the Toffoli-CNOT-based one \cite{Fu-T}.  Moreover, extra ancillary photon is necessary for the schemes presented in Refs. \cite{Bugu,Fu-T}, and is not required in our scheme.  Remarkably, our protocol is less complex with a higher success probability than any of the other protocols for the generation of a larger $W$ state with the same size \cite{Ozdemir,Bugu,Fu-T}.

\begin{table}[htb]
\centering \caption{Quantum resource required and success probability of various protocols for creating a larger $W$ state. $H$ is an extra ancillary $H$-polarized photon required for the fusion program.}
\begin{tabular}{cccccc}
\toprule
 Proposed                           &   Initial                  &       Success  &      Success         &  Recycle                 & Fail          \\
protocol                           &    state                   &        result  &      probability     &  probability             & probability  \\
\midrule
with $I$ \cite{Ozdemir}             &   $W_{m}, W_{n} $          &  $W_{m+n-2}$   &  $\frac{m+n-2}{mn}$  &  $\frac{(m-1)(n-1)}{mn}$ &  $\frac{1}{mn}$ \\
with 1 Fredkin \cite{Bugu}          &   $W_{m}, W_{n}, H $       &  $W_{m+n-1}$   &  $\frac{m+n-1}{mn}$  &  $\frac{(m-1)(n-1)}{mn}$ &  0              \\
with 1 Toffoli, 1 CNOT \cite{Fu-T}  &   $W_{m}, W_{n}, H $       &  $W_{m+n-1}$   &  $\frac{m+n-1}{mn}$  &  $\frac{(m-1)(n-1)}{mn}$ &  0              \\
ours with 1 partial-swap            &   $W_{m}, W_{n} $          &  $W_{m+n-1}$   &  $\frac{m+n-1}{mn}$  &  $\frac{(m-1)(n-1)}{mn}$ &  0  \\
\bottomrule
\end{tabular}\label{table1}
\end{table}

\subsection{Fusing $n$ arbitrary-sized $|W_m\rangle$ states into a large-scalable $|W_{nm-n+1}\rangle$ state}

\begin{figure}[!h]    
\begin{center}
\includegraphics[width=9 cm,angle=0]{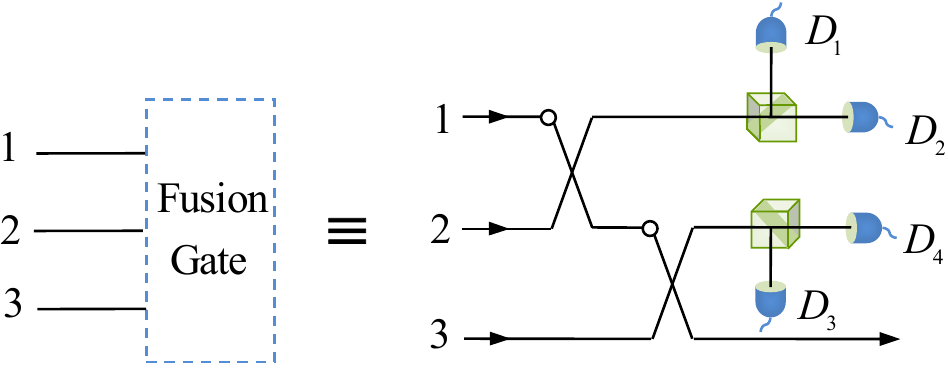}
\caption{Fusion gate for fusing three $W$ states of arbitrary size to obtain a larger $W$ state.} \label{w3}
\end{center}
\end{figure}

Fig. \ref{w3} displays a scheme for fusing $|W_n\rangle_A$, $|W_m\rangle_B$, and $|W_t\rangle_C$ states into $|W_{n+m+t-2}\rangle$ by using two partial-swap gates.  We denote polarization-based entangled $W$ states of Alice, Bob, and Charlie as
\begin{eqnarray}              \label{eq7}
\begin{split}
|W_n\rangle_A=(|(n-1)_H\rangle_a|1_V\rangle_1+\sqrt{n-1}|W_{n-1}\rangle_a|1_H\rangle_1)/\sqrt{n},
\end{split}
\end{eqnarray}
\begin{eqnarray}              \label{eq8}
\begin{split}
|W_m\rangle_B=(|(m-1)_H\rangle_b|1_V\rangle_2+\sqrt{m-1}|W_{m-1}\rangle_b|1_H\rangle_2)/\sqrt{m},
\end{split}
\end{eqnarray}
\begin{eqnarray}              \label{eq9}
\begin{split}
|W_t\rangle_C=(|(t-1)_H\rangle_c|1_V\rangle_3+\sqrt{t-1}|W_{t-1}\rangle_c|1_H\rangle_3)/\sqrt{t}.
\end{split}
\end{eqnarray}

As shown  Fig. \ref{w3}, after Alice, Bob, and Charlie send one of their photons ($|1_H\rangle_1$ and $|1_V\rangle_1$, $|1_H\rangle_2$ and $|1_V\rangle_2$, $|1_H\rangle_3$ and $|1_V\rangle_3$) to the two partial-swap gates through modes 1, 2, and 3, respectively, the two partial-swap gates lead to the following transformations:
\begin{eqnarray}              \label{eq10}
\begin{split}
|W_n\rangle \otimes &|W_m\rangle \otimes |W_t\rangle\\\rightarrow&
\frac{1}{\sqrt{nmt}}                     |(n-1)_H\rangle_a|(m-1)_H\rangle_b|(t-1)_H\rangle_c|1_V\rangle_1|1_V\rangle_2|1_V\rangle_3\\
&+\frac{\sqrt{n-1}}{\sqrt{nmt}}            |W_{n-1}\rangle_a|(m-1)_H\rangle_b|(t-1)_H\rangle_c|1_V\rangle_1|1_V\rangle_2|1_H\rangle_3\\
&+\frac{\sqrt{m-1}}{\sqrt{nmt}}            |(n-1)_H\rangle_a|W_{m-1}\rangle_b|(t-1)_H\rangle_c|1_V\rangle_1|1_V\rangle_2|1_H\rangle_3\\
&+\frac{\sqrt{t-1}}{\sqrt{nmt}}            |(n-1)_H\rangle_a|(m-1)_H\rangle_b|W_{t-1}\rangle_c|1_V\rangle_1|1_V\rangle_2|1_H\rangle_3\\
&+\frac{\sqrt{(n-1)(m-1)}}{\sqrt{nmt}}     |W_{n-1}\rangle_a|W_{m-1}\rangle_b|(t-1)_H\rangle_c|1_H\rangle_1|1_V\rangle_2|1_H\rangle_3\\
&+\frac{\sqrt{(n-1)(t-1)}}{\sqrt{nmt}}     |W_{n-1}\rangle_a|(m-1)_H\rangle_b|W_{t-1}\rangle_c|1_V\rangle_1|1_H\rangle_2|1_H\rangle_3\\
&+\frac{\sqrt{(m-1)(t-1)}}{\sqrt{nmt}}     |(n-1)_H\rangle_a|W_{m-1}\rangle_b|W_{t-1}\rangle_c|1_V\rangle_1|1_H\rangle_2|1_H\rangle_3\\
&+\frac{\sqrt{(n-1)(m-1)(t-1)}}{\sqrt{nmt}}|W_{n-1}\rangle_a
|W_{m-1}\rangle_b|W_{t-1}\rangle_c|1_H\rangle_1|1_H\rangle_2|1_H\rangle_3.
\end{split}
\end{eqnarray}

Eq. (\ref{eq10}) implies the following four possible outcomes:

(i) When the photon in mode 1 is $V$-polarized state,  the photon in mode 2 is also $V$-polarized state and detectors $D_1$ and $D_3$ click,  the system collapses into the successful  state $|W_{n+m+t-2}\rangle$
\begin{eqnarray}              \label{eq11}
\begin{split}
|W_{n+m+t-2}\rangle=
&\frac{1}{\sqrt{nmt}}                      |(n-1)_H\rangle_a|(m-1)_H\rangle_b|(t-1)_H\rangle_c|1_V\rangle_3\\
&+\frac{\sqrt{n-1}}{\sqrt{nmt}}            |W_{n-1}\rangle_a|(m-1)_H\rangle_b|(t-1)_H\rangle_c|1_H\rangle_3\\
&+\frac{\sqrt{m-1}}{\sqrt{nmt}}            |(n-1)_H\rangle_a|W_{m-1}\rangle_b|(t-1)_H\rangle_c|1_H\rangle_3\\
&+\frac{\sqrt{t-1}}{\sqrt{nmt}}            |(n-1)_H\rangle_a|(m-1)_H\rangle_b|W_{t-1}\rangle_c|1_H\rangle_3.
\end{split}
\end{eqnarray}

(ii) When the photon in mode 1 is $H$-polarized state, the photon in mode 2 is also $H$-polarized state and detectors $D_2$ and $D_4$ click,  the system collapses into the recyclable state
\begin{eqnarray}             \label{eq12}
\frac{\sqrt{(n-1)(m-1)(t-1)}}{\sqrt{nmt}}|W_{n-1}\rangle_a|W_{m-1}\rangle_b|W_{t-1}\rangle_c|1_H\rangle_3.
\end{eqnarray}

(iii) When the photon in mode 1 is $H$-polarized state, the photon in mode 2 is $V$-polarized state and detectors $D_2$ and $D_3$ click,  the system  collapses into the ``garbage'' state
\begin{eqnarray}             \label{eq13}
\frac{\sqrt{(n-1)(m-1)}}{\sqrt{nmt}}|W_{n-1}\rangle_a|W_{m-1}\rangle_b|(t-1)_H\rangle_c|1_H\rangle_3.
\end{eqnarray}
We call this case ``a partial recyclable''  because the states between Alice and Bob remain a $W$-state but Charlie needs to prepare a new $W$ state for the subsequent round of the fusion process.

(iv) When the photon in mode 1 is $V$-polarized state, the photon in mode 2 is $H$-polarized state and detectors $D_1$ and $D_4$ click,  the system collapses into the ``garbage'' state
\begin{eqnarray}             \label{eq14}
\begin{split}
\frac{\sqrt{(t-1)(m+n-2)}}{\sqrt{nmt}}|W_{n+m-2}\rangle_{a,b}|W_{t-1}\rangle_c|1_H\rangle_3.
\end{split}
\end{eqnarray}
We call this case ``a partial success''  because the state between Alice and Bob has been fused but not Charlie.

\begin{figure}[!h]    
\begin{center}
\includegraphics[width=12 cm,angle=0]{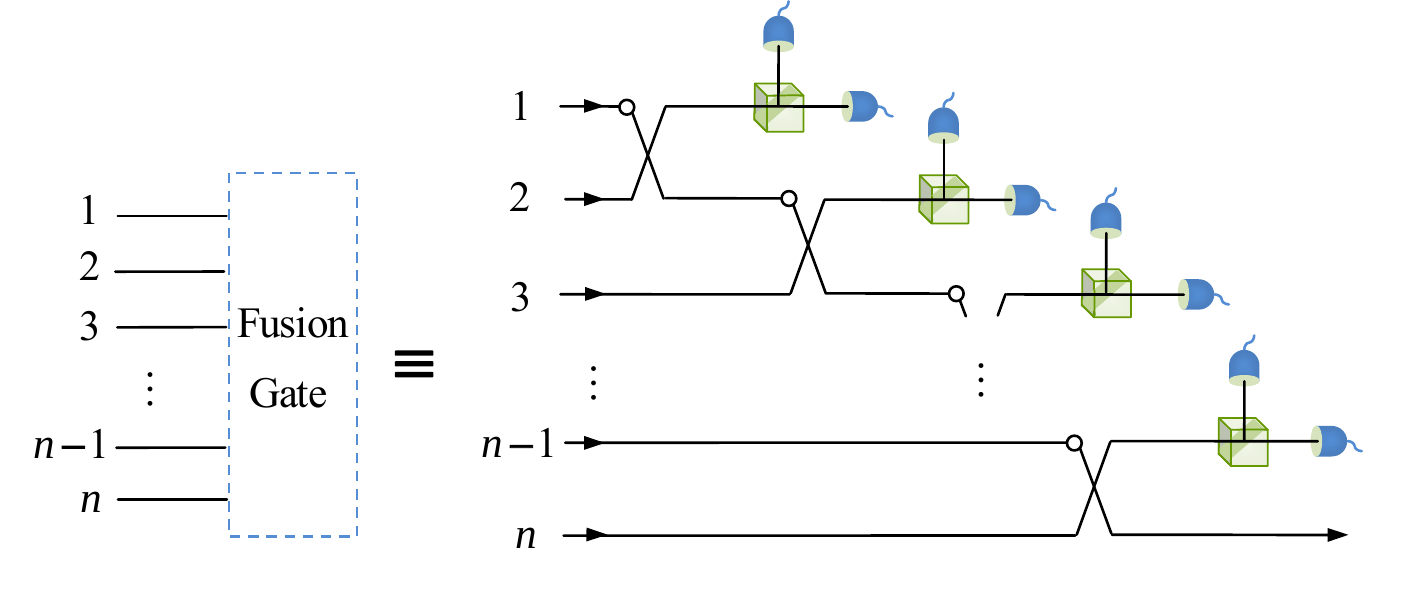}
\caption{Fusion gate for fusing $n$ arbitrary-sized $W$ states simultaneously.} \label{wn}
\end{center}
\end{figure}

Fig. \ref{wn} shows a scheme for fusing multiple $W$ states of arbitrary size simultaneously.
Table \ref{table2} compares the success and failure probabilities and an estimation of the required quantum resources for our proposal against previous schemes.
Compared with other proposals for generating a $W$ state of given size, our proposal scores a higher success probability and a lower failure probability, with a simpler network.

\begin{table}[htb]
\centering \caption{Quantum resources required and success probabilities of various protocols for fusing multiple $W$ states into a larger one simultaneously. $F=((5-m-n-z-t)-(n-1)(m-1)(t-1)(z-1))/(mntz)$.}
\begin{tabular}{cccccc}
\toprule
 Proposed                          &  Initial         &       Achieved    &  Success                  &   Fail                          \\
protocol                          &   state          &        state     &  probability              &   probability                     \\
\midrule
with 1 Fredkin \cite{Bugu}         &  $W_m,W_n,W_t,H$ &  $W_{m+n+t-3}$    &  $\frac{m+n+t-3}{mnt}$    &   $\frac{(t-1)(m+n-2)+1}{mnt}$  \\

ours with 2 partial-swaps          & $W_m,W_n,W_t,$      &  $W_{m+n+t-2}$    &  $\frac{m+n+t-2}{mnt}$    &   0 \\

with 1 Toffoli, 3 CNOTs \cite{Yesilyurt}  &  $W_m,W_n,W_t,W_z$   &  $W_{m+n+t+z-4}$  &  $\frac{m+n+t+z-4}{mntz}$ &   $F$ \\

ours with 3 partial-swaps          &  $W_m,W_n,W_t,W_z$   &  $W_{m+n+t+z-3}$  &  $\frac{m+n+z+t-3}{mntz}$ &  0    \\

\bottomrule
\end{tabular}\label{table2}
\end{table}

\section{Linear-optics fusion-based $W$ state using auxiliary spatial degrees of freedom}\label{sec3}

\begin{figure}[!h]    
\begin{center}
\includegraphics[width=8 cm,angle=0]{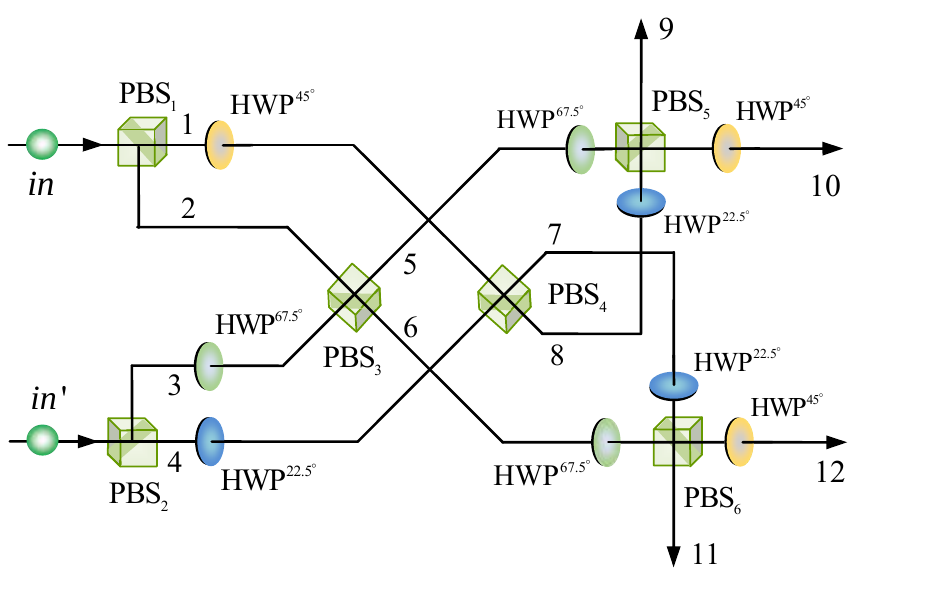}
\caption{Linear-optical post-selected partial-swap gate.
HWP$^{45^\circ}$ is a half-wave plate (HWP) rotated by 45$^\circ$ to induce the transformation $|H\rangle \leftrightarrow|V\rangle$.
Setting HWP$^{22.5^\circ}$ (HWP$^{67.5^\circ}$) to 22.5$^\circ$ (67.5$^\circ$) completes
$|H\rangle\leftrightarrow(|H\rangle+|V\rangle)/\sqrt{2}$ and $|V\rangle\leftrightarrow(|H\rangle-|V\rangle)/\sqrt{2}$ ($|H\rangle\leftrightarrow(-|H\rangle+|V\rangle)/\sqrt{2}$ and $|V\rangle\leftrightarrow(|H\rangle+|V\rangle)/\sqrt{2}$).} \label{p-swap}
\end{center}
\end{figure}

Based on Sec. \ref{sec2}, one can see that the key component of our fusion gates is the partial-swap gate described by  Eq. (\ref{eq4}). The matrix form of this partial-swap gate in the basis $\{|1_H\rangle|1_H\rangle$, $|1_H\rangle|1_V\rangle$, $|1_V\rangle|1_H\rangle$, $|1_V\rangle|1_V\rangle\}$ can be written as
\begin{eqnarray}             \label{eq15}
N_{\text{p-swap}}=\left(
  \begin{array}{cccc}
    1 & 0 & 0 & 0 \\
    0 & 0 & 0 & 0 \\
    0 & 1 & 1 & 0 \\
    0 & 0 & 0 & 1 \\
  \end{array}
\right).
\end{eqnarray}
Obviously, this operation is not a unitary one due to $NN^\dagger \neq N^\dagger N \neq I$, with $I$ being an identity matrix.

The nonunitary gate can be implemented by utilizing the framework of quantum measurement, or by expanding the state space to a larger one, and then performing a proper unitary operation and an orthogonal measurement in the enlarged space in succession.  Here, we employ auxiliary spatial degrees of freedom introduced by polarizing beam splitters PBS$_1$  and PBS$_2$ (see Fig. \ref{p-swap}) to implement the nonunitary polarization partial-swap gate. Next, we provide a step-by-step description of our protocol for implementing this partial-swap gate.

We consider photons 1 and 2 as being initially prepared in an arbitrary two-qubit polarization-encoded state
\begin{eqnarray}              \label{eq16}
\begin{split}
|\psi_0\rangle=&\alpha_1 |1_H\rangle_{in}|1_H\rangle_{in'} + \alpha_2 |1_H\rangle_{in}|1_V\rangle_{in'} \\
&+\alpha_3 |1_V\rangle_{in}|1_H\rangle_{in'} + \alpha_4 |1_V\rangle_{in}|1_V\rangle_{in'}.
\end{split}
\end{eqnarray}


In the first step,  as shown in Fig. \ref{p-swap}, photons 1 and 2 pass through PBS$_1$  and PBS$_2$, respectively.
Next, photons in modes 1, 3, and 4 interact with half-wave plates (HWP) oriented at $45^\circ$ (HWP$^{45^\circ}$), $67.5^\circ$ (HWP$^{67.5^\circ}$), and $22.5^\circ$ (HWP$^{22.5^\circ}$), respectively.  The PBSs and HWPs cause the state to evolve from $|\psi\rangle_0$ to
\begin{eqnarray}              \label{eq17}
\begin{split}
|\psi_1\rangle=&\frac{1}{\sqrt{2}}(
  \alpha_1 |1_V\rangle_1(|1_H\rangle_4+|1_V\rangle_4)
 +\alpha_2 |1_V\rangle_1(|1_H\rangle_3+|1_V\rangle_3)\\
&+\alpha_3 |1_V\rangle_2(|1_H\rangle_4+|1_V\rangle_4)
 +\alpha_4 |1_V\rangle_2(|1_H\rangle_3+|1_V\rangle_3)).
\end{split}
\end{eqnarray}
%
%
A PBS  transmits the $H$-polarized and reflects the $V$-polarized components. Therefore, PBS$_1$ and PBS$_2$ impart the spatial  degrees of freedom of the incident photon.  The
HWPs oriented at $45^\circ$ (denoted HWP$^{45^\circ}$) induce the qubit-flip operation $|1_H\rangle \leftrightarrow |1_V\rangle$
while the HWP$^{67.5^\circ}$ results in
\begin{eqnarray}                  \label{eq18}
\begin{split}
&|1_H\rangle \leftrightarrow\frac{1}{\sqrt{2}}(-|1_H\rangle+|1_V\rangle), \quad
|1_V\rangle\leftrightarrow\frac{1}{\sqrt{2}}(|1_H\rangle+|1_V\rangle).
\end{split}
\end{eqnarray}
Finally, the HWP$^{22.5^\circ}$  completes the transformation
\begin{eqnarray}                  \label{eq19}
\begin{split}
&|1_H\rangle\leftrightarrow\frac{1}{\sqrt{2}}(|1_H\rangle+|1_V\rangle), \qquad
|1_V\rangle\leftrightarrow\frac{1}{\sqrt{2}}(|1_H\rangle-|1_V\rangle).
\end{split}
\end{eqnarray}

In the second step, the photons in modes 2 and 3 are  then mixed at PBS$_3$ before going through HWP$^{67.5^\circ}$ while the photons in modes 1 and 4 are mixed at PBS$_4$ before going through HWP$^{22.5^\circ}$. The completion of these operations leads to the joint state
\begin{eqnarray}              \label{eq20}
\begin{split}
|\psi_2\rangle=&\frac{1}{2\sqrt{2}}(
   \alpha_1 (|1_H\rangle_7-|1_V\rangle_7)(|1_H\rangle_7+|1_V\rangle_7+|1_H\rangle_8-|1_V\rangle_8)\\
& +\alpha_2 (|1_H\rangle_7-|1_V\rangle_7)(-|1_H\rangle_5+|1_V\rangle_5+|1_H\rangle_6+|1_V\rangle_6)\\
&+\alpha_3  (|1_H\rangle_5+|1_V\rangle_5)(|1_H\rangle_7+|1_V\rangle_7+|1_H\rangle_8-|1_V\rangle_8)\\
& +\alpha_4 (|1_H\rangle_5+|1_V\rangle_5)(-|1_H\rangle_5+|1_V\rangle_5+|1_H\rangle_6+|1_V\rangle_6)).
\end{split}
\end{eqnarray}

In the third step, the photons in modes 5 and 8 (6 and 7) are combined at PBS$_5$ (PBS$_6$), and the photon in mode 10 (12) passes through HWP$^{45^\circ}$ (HWP$^{45^\circ}$). The operations  $\text{PBS}_5 \rightarrow \text{HWP}^{45^\circ}$ and $\text{PBS}_6 \rightarrow \text{HWP}^{45^\circ}$ make $|\psi\rangle_2$ become
\begin{eqnarray}              \label{eq21}
\begin{split}
|\psi_3\rangle=&\frac{1}{2\sqrt{2}}(
  \alpha_1  (|1_H\rangle_{11}-|1_H\rangle_{12}) (|1_H\rangle_{11}+|1_H\rangle_{12}+|1_H\rangle_{9}-|1_H\rangle_{10})\\
& +\alpha_2 (|1_H\rangle_{11}-|1_H\rangle_{12})(-|1_V\rangle_{10}+|1_V\rangle_{9}+|1_V\rangle_{12}+|1_V\rangle_{11})\\
&+\alpha_3  (|1_V\rangle_{10}+|1_V\rangle_9)    (|1_H\rangle_{11}+|1_H\rangle_{12}+|1_H\rangle_{9}-|1_H\rangle_{10})\\
& +\alpha_4 (|1_V\rangle_{10}+|1_V\rangle_9)   (-|1_V\rangle_{10}+|1_V\rangle_{9}+|1_V\rangle_{12}+|1_V\rangle_{11})).
\end{split}
\end{eqnarray}

Eq. (\ref{eq21}) indicates that the two-qubit partial-swap operation (i.e., exchanges the information of the two photons, conditional on the first photon being $H$-polarized) is completed when a coincidence is observed  between modes 9 and 11 (10 and 12).   Table \ref{table3}  lists the photon count rates in modes 9 and 11 (10 and 12) for computing basis inputs.

\begin{table}[htb]
\centering \caption{Coincident expectation values calculated for the four logic basis inputs. }
\begin{tabular}{cccccccccccccccc}
\toprule
 & $\langle n_{|1_H\rangle_9}n_{|1_H\rangle_{11}}\rangle$ & $\langle n_{|1_H\rangle_9}n_{|1_V\rangle_{11}}\rangle$& $\langle n_{|1_V\rangle_9}n_{|1_H\rangle_{11}}\rangle$     & $\langle n_{|1_V\rangle_9}n_{|1_V\rangle_{11}}\rangle$\\

\cline{2-5}

Input  & $\langle n_{|1_H\rangle_{10}}n_{|1_H\rangle_{12}}\rangle$ & $\langle n_{|1_H\rangle_{10}}n_{|1_V\rangle_{12}}\rangle$ & $\langle n_{|1_V\rangle_{10}}n_{|1_H\rangle_{12}}\rangle$ & $\langle n_{|1_V\rangle_{10}}n_{|1_V\rangle_{12}}\rangle$ \\

\midrule

$|1_H\rangle_{in}|1_H\rangle_{in'}$   & 1/8  &  0  &  0    & 0     \\

$|1_H\rangle_{in}|1_V\rangle_{in'}$   &  0   &  0  &  1/8  & 0     \\

$|1_V\rangle_{in}|1_H\rangle_{in'}$  & 0    &  0  &  1/8  & 0     \\

$|1_V\rangle_{in}|1_V\rangle_{in'}$   & 0    &  0  &  0    & 1/8   \\

\bottomrule
\end{tabular}\label{table3}
\end{table}

\section{Discussion and Conclusion}\label{sec4}

In this paper, we have proposed an effective scheme for fusing $|W_n\rangle$ and $|W_m\rangle$ states into a large-size $|W_{n+m-1}\rangle$ state by using a partial-swap gate. We have also designed a scheme for fusing multiple $W$ states of arbitrary size simultaneously (see Fig. \ref{wn}). By exploiting the spatial  degrees of freedom of single-photons introduced by the PBSs, the partial-swap gate was implemented using an optically polarized architecture designed with linear-optical elements.

As shown in Table \ref{table1}, our scheme outperforms  previous ones in fusing two $W$ states of arbitrary size into a large-sized $W$ state. An ancillary photon, which is necessary in the Fredkin- and Toffoli-based schemes  \cite{Bugu,Fu-T} to create  a $|W_{n+m-1}\rangle$ state, is not required in our scheme. Moreover, our scheme minimizes failure outcomes. From Table \ref{table2}, one can see that, if the gate (Fredkin, Toffoli, CNOT, and partial-swap gates) operations are considered, the fail probability  in the presented scheme is lower than that with the Fredkin- and Toffoli-based schemes \cite{Bugu,Yesilyurt}.

Our presented scheme has the further advantage of reducing cost in terms of the number of two-qubit gates. In previous studies \cite{Bugu,Fu-T}, the fusion of two $W$ states required either one Fredkin gate, or one Toffoli and one CNOT gate. Our presented approach requires just one partial-swap gate. Notably, the optimal cost of an unconstructed Toffoli or Fredkin gate is five two-qubit gates \cite{Optimal1,Optimal2}. If we impose a further condition of using only CNOT gates, at least six CNOT gates are required to synthesize a Toffoli and at least eight for a Fredkin gate \cite{Optimal3}. In contrast, the current proposals based on  partial-swap gates surpass the Fredkin-gate scheme \cite{Bugu} and Toffoli-CNOT-scheme \cite{Fu-T}, and also surpass the scheme based on one Toffoli gate and three CNOT gates \cite{Yesilyurt}  (see Table \ref{table1} and Table \ref{table2}).

Another important advantage of the our scheme is its increased success probability. It is known that entangling quantum gates can be implemented only probabilistically using linear-optical elements. With a polynomial quantum resource, a linear-optical CNOT gate can be implemented  with a maximal success probability of 3/4 \cite{KLM}, and a post-selected CNOT gate with a success probability of 1/9 \cite{1/9}. The most efficient scheme for a CNOT gate with a success probability of 1/4  is achieved with the help of a maximally entangled photon pair  \cite{1/41,1/42}.  Moreover, the ideal  success probability of a Toffoli gate is 1/9 under a linear-optics setup \cite{Toffoli}.  At present, the optimal success probability of a linear optical Fredkin gate is 1/64 \cite{Fredkin}. In contrast, the proposed partial-swap gate with a success probability of 1/4 is achievable. To sum up, our partial-swap-based fusion schemes  outperform the Fredkin-based \cite{Bugu} and Toffoli-CNOT-based schemes \cite{Yesilyurt,Fu-T} in term of the cost and success probability.

\section*{Acknowledgment}

The work was supported by the National Natural Science Foundation of China under Grant No. 11604012, and the Fundamental Research Funds for the Central Universities under Grant Nos.  FRF-BR-17-004B and 230201506500024, and a grant from the China Scholarship Council. KLC is supported by the Ministry of Education and the National Reserach Foundation Singapore.



\end{document}